\documentclass[12pt]{article}
\begin{document}

\begin{titlepage}

\hspace*{\fill}\parbox[t]{5cm}
{WORKING PAPER  \\

December 2010 \\

Financial Analysts Journal \\ Volume 67, No.~4 \\ July/August 2011, p.~42-49
}
\vskip2cm
\begin{center}
{\Large \bf Diversification Return, Portfolio Rebalancing,\\
\bigskip and the Commodity Return Puzzle}
\medskip

\bigskip\bigskip\bigskip

\medskip
\bigskip\bigskip\bigskip\bigskip
{\large {\bf Scott Willenbrock}\footnote{Registered Investment Advisor}} \\
\bigskip\bigskip\medskip
Department of Physics, University of Illinois at Urbana-Champaign \\
\bigskip 1110 West Green Street, Urbana, IL\ \ 61801 \\ \bigskip
\end{center}

\bigskip\bigskip\bigskip

\begin{abstract}
Diversification return is an incremental return earned by a rebalanced portfolio of assets.  The diversification return of a rebalanced portfolio is often incorrectly ascribed to a reduction in variance.  We argue that the underlying source of the diversification return is the rebalancing, which forces the investor to sell assets that have appreciated in relative value and buy assets that have declined in relative value, as measured by their weights in the portfolio.
In contrast, the incremental return of a buy-and-hold portfolio is driven by the fact that the assets that perform the best become a greater fraction of the portfolio.  We use these results to resolve two puzzles associated with the Gorton and Rouwenhorst index of commodity futures, and thereby obtain a clear understanding of the source of the return of that index.  Diversification return can be a significant source of return for any rebalanced portfolio of volatile assets.
\end{abstract}

\end{titlepage}

\section{Introduction}\label{sec:intro}

The term ``diversification return'' was coined by Booth and Fama (1992) in the context of a rebalanced portfolio, that is, a portfolio with a constant percentage invested in each asset.  They showed that the contribution of each asset to the portfolio compound return, which they dubbed the return contribution, exceeds the asset's compound return by an incremental amount that they named the diversification return.  The portfolio's diversification return is the weighted average of the assets' diversification returns.

Since Booth and Fama (1992) consider a rebalanced portfolio throughout their work, one is led to wonder if there are two separate, perhaps related, aspects to the incremental return: diversification and rebalancing.  Several authors, including Erb and Harvey (2006), Gorton and Rouwenhorst (2006b), and Idzorek (2006,2007) have argued in favor of this point of view, to different degrees.  For example, Gorton and Rouwenhorst (2006b) regard the diversification return of a portfolio as the difference between the portfolio geometric average return and the (weighted) geometric average returns of the individual assets, regardless of whether the weights are constant or not.

In this paper we revisit the issue of diversification return and portfolio rebalancing.  We show that diversification return can be precisely defined in the context of a rebalanced portfolio.  We argue that the reduction in variance inherent in a diversified portfolio is a necessary, but not sufficient, condition in order to earn a diversification return, in contrast with the argument of Gorton and Rouwenhorst (2006b).

We clarify the underlying source of diversification return.  Booth and Fama (1992) derive the diversification return in terms of the difference between the variance of an asset and its covariance with the portfolio.  While this is an elegant and useful approach, it masks the fact that the diversification return stems from selling assets that have appreciated in relative value and buying assets that have declined in relative value, as measured by their weights in the portfolio.

We also analyze a buy-and-hold portfolio.  While there is no diversification return for such a portfolio, it is possible for the portfolio to have an incremental return relative to the initially-weighted average of the compound returns of the assets.  This stems from the fact that, over time, a buy-and-hold portfolio will increase the weights of the assets that perform the best.  However, this also changes the risk profile of the portfolio.  In contrast, an investor earns a diversification return in a rebalanced portfolio while maintaining a constant risk profile.

Finally, we use these results to resolve two aspects of the commodity return puzzle stated by Idzorek (2006,2007).  We argue that the excess return (above the risk-free return) of the Gorton and Rouwenhorst (2006a) commodity futures index, which is rebalanced monthly, can be largely accounted for by the diversification return, as argued by Erb and Harvey (2006).  However, if the index is not rebalanced, it generates a significant incremental excess return as a buy-and-hold portfolio, because the compound returns of the underlying commodity futures in the index have a wide range of values.
Thus there is no contradiction; the Gorton and Rouwenhorst commodity futures index generates an excess return in both its rebalanced and unrebalanced incarnations, but for totally different and unrelated reasons.

\newpage

The paper is organized as follows.  In Section~\ref{sec:geometric} we present a simple derivation of an approximate formula for the diversification return of a rebalanced portfolio, highlighting that the arithmetic average return of an asset is a useful intermediate quantity only if the asset weights are held constant.  In Section~\ref{sec:source} we identify the source of the diversification return of a rebalanced portfolio, and provide a precise definition of the diversification return.  We also discuss several examples.  In Section~\ref{sec:buy} we argue that a buy-and-hold portfolio does not have a diversification return, but that it may still have an incremental return from a different source.  In Section~\ref{sec:commodity} we apply these results to solve two aspects of the commodity return puzzle.  We summarize our results in Section~\ref{sec:conclusions}, and discuss the relevance of these findings to investors.

\section{Geometric and arithmetic average returns}\label{sec:geometric}

The beginning point of our discussion is the relationship between the arithmetic average return and the geometric average return, valid for an individual asset as well as for a portfolio.  Following Booth and Fama (1992), we derive in the Appendix the approximate relation
\begin{equation}
g \approx \bar r - \frac{1}{2}\sigma^2
\label{eq:geometric}
\end{equation}
where $g$ is the geometric average return, $\bar r$ is the arithmetic average of the simple returns $r$, and $\sigma^2$ is the variance of the simple returns.  While Booth and Fama (1992) framed the discussion in terms of compound return, we will find it advantageous to work with the geometric average return, as in Erb and Harvey (2006) and Gorton and Rouwenhorst (2006b).

Equation (\ref{eq:geometric}) is simple, but in general it is not useful.  Let us recall that the arithmetic average return is a misleading measure of an asset's or portfolio's performance.  An asset or portfolio with simple returns of $50\%$ and $-50\%$ has an arithmetic average return of zero, yet the investor has a loss of $25\%$.  In contrast, the geometric average return is a precise measure of the return of an asset or portfolio. Given an asset's or portfolio's geometric average return over $T$ holding periods, the ratio of the final value of the asset or portfolio to its initial value is $(1+g)^T$.  No such relation exists for the arithmetic average return; it is impossible to calculate the relation between final and initial asset or portfolio values based on the arithmetic average return alone.  Equation (\ref{eq:geometric}) may be regarded as a reminder of the dependence of the arithmetic average return on the volatility of the asset or portfolio.

Given the misleading nature of the arithmetic average return, it cannot in general be used by itself to judge an asset's or portfolio's performance.  There is an exception to this statement, and that is for a rebalanced portfolio, that is, a portfolio that rebalances to constant weights at the end of each holding period.  To see this, consider the relation between the simple return of a portfolio, $r_p$, and the simple returns of the assets, $r_i$,
\begin{equation}
r_p=\sum_i w_i r_i
\label{eq:simple}
\end{equation}
where $w_i$ is the weight of the asset in the portfolio ($\sum_i w_i = 1$).  If we take the average of this equation, we find a useful relation between the arithmetic average return of the portfolio, $\bar r_p$, and the arithmetic average returns of the assets, $\bar r_i$,
\begin{equation}
\bar r_p=\sum_i w_i \bar r_i
\label{eq:arithmetic}
\end{equation}
{\it if and only if the weights, $w_i$, are held constant}.  This is the key relation that allows us to derive the diversification return.  Using Eq.~(\ref{eq:geometric}) on both sides of this equation, we obtain
\begin{equation}
g_p + \frac{1}{2}\sigma^2_p\approx\sum_i w_i \left(g_i + \frac{1}{2}\sigma^2_i\right)\;.
\label{eq:sigmap}
\end{equation}
We now use the relation between a portfolio's variance and the covariance of each asset's simple returns with the portfolio's simple returns, $\sigma_{ip}^2$,
\begin{equation}
\sum_i w_i \sigma_{ip}^2 = \sigma_p^2
\end{equation}
which, like Eq.~(\ref{eq:arithmetic}), is only true if the weights are held constant.  This yields the desired result
\begin{equation}
g_p \approx \sum_i w_i \left(g_i + \frac{1}{2}(\sigma^2_i-\sigma_{ip}^2)\right)\;.
\label{eq:totalreturn}
\end{equation}
The diversification return of an asset is thus approximately half the difference between the asset's variance and the asset's covariance with the portfolio.  The diversification return of the portfolio is the weighted average of the diversification returns of the assets,
\begin{equation}
{\rm Diversification\ Return} \approx \frac{1}{2}\sum_i w_i (\sigma^2_i-\sigma_{ip}^2)\;.
\label{eq:return}
\end{equation}

Let us underscore once again that it is essential to maintain (nearly) constant weights in order to obtain a diversification return. It is only in this case that the arithmetic average return, a generally misleading measure of an asset's or portfolio's performance, becomes a useful intermediate quantity via Eq.~(\ref{eq:arithmetic}).  In more general settings, one cannot infer anything directly from Eq.~(\ref{eq:geometric}).  It is faulty logic to claim that a portfolio's geometric average return is increased solely by the decrease in variance that is inherent in a diversified portfolio, as argued by Gorton and Rouwenhorst (2006b).

\section{The source of diversification return}\label{sec:source}

The result for the diversification return, Eq.~(\ref{eq:return}), is both elegant and useful.  Although it applies only to a rebalanced portfolio, that is, a portfolio with a constant percentage invested in each asset, it does not depend on the details of the rebalancing.  All that matters is the variances of the assets and their covariances with the portfolio.

Nevertheless, the underlying source of the diversification return is contained in the rebalancing.  Rebalancing a portfolio involves selling assets that have appreciated in relative value and buying assets that have declined in relative value, as measured by their weights in the portfolio.  This contrarian activity generates incremental returns as the assets fluctuate in value.

Consider a portfolio in which each asset has vanishing geometric average return, that is, has the same value as at the beginning of the $T$ holding periods.  If the portfolio is not rebalanced, it has a geometric average return of zero.  However, rebalancing the portfolio will generate incremental returns.  Thus the entire return of the portfolio is generated by rebalancing.  This is the diversification return.

Now consider a portfolio in which each asset has the identical, nonvanishing, geometric average return.  If the portfolio is not rebalanced, it has the same geometric average return as its constituents.  Rebalancing the portfolio will generate incremental returns.  Thus the return of the portfolio is the sum of the geometric average return of the assets and an incremental return due to rebalancing. Comparing with Eq.~(\ref{eq:totalreturn}), the first term on the right-hand side is the geometric average return of the assets.  Thus the second term, which is the diversification return, is the incremental return due to rebalancing.

Finally, consider the general case of a portfolio in which each asset has a different nonvanishing geometric average return.  The geometric average return of the portfolio is given approximately by Eq.~(\ref{eq:totalreturn}).  The first term depends only on the geometric average returns of the assets that compose the portfolio, and not at all on the volatility of the assets.  The second term, which is the diversification return, depends on the volatility of the assets.  If the assets had zero volatility, that is, if their variances vanished, then their covariances with the portfolio would also vanish, and the diversification return would be zero.  Hence the diversification return is driven by the volatility of the assets.  The underlying source of the diversification return is thus the selling of assets that have appreciated in relative value and the buying of assets that have declined in relative value.  An alternative name for diversification return might be volatility return.

A simple example demonstrates the process.  Consider a portfolio consisting of two assets that are initially equally weighted.  The simple returns of the first asset are $+25\%,-20\%$, while the simple returns of the second asset are the reverse, $-20\%,+25\%$.  Each asset has a vanishing geometric average return over the two holding periods.  However, if the portfolio is rebalanced to equal weights after the first holding period, the portfolio has a gain of $5.06\%$, corresponding to a geometric average return of $2.5\%$.  This is a diversification return.

The separation of the portfolio geometric average return into two separate pieces is approximate, simply because Eq.~(\ref{eq:totalreturn}) is approximate.  However, the above discussion suggests a precise definition of the diversification return of a rebalanced portfolio.  The diversification return is the difference between the geometric average returns of a rebalanced portfolio of volatile assets, and a rebalanced portfolio of hypothetical assets with the same weights and geometric average returns as the true assets, but zero volatility.  An appropriate name for the geometric average return of the hypothetical portfolio might be the strategic return, because it depends only on the geometric average returns of the assets and their weights in the portfolio.  Thus, for a rebalanced portfolio,
\begin{equation}
{\rm Diversification\ return} \equiv {\rm Geometric\ average\ return} - {\rm Strategic\ return}\;.
\label{eq:diversification}
\end{equation}

If an asset has zero volatility, then its geometric average return is identical to its arithmetic average return. Thus, using Eq.~(\ref{eq:arithmetic}), we see that the strategic return of a portfolio is given exactly by
\begin{equation}
{\rm Strategic\ return} \equiv \sum_i w_i g_i\;.
\label{eq:strategic}
\end{equation}
It is this equation that motivates our use of the geometric average return instead of the compound return used by Booth and Fama (1992).  The diversification return is given exactly by Eq.~(\ref{eq:diversification}) and
approximately by Eq.~(\ref{eq:return}).

A detailed example of strategic return and diversification return in the context of a simple portfolio is given in Table~\ref{tab:portfolio}.  The portfolio has 50\% invested in the S\&P 500 Index (Total Return) and 50\% invested in the Barclays US Long Treasury Index (Total Return) on January 1, 2000.  The portfolio is rebalanced at the end of each year.  The geometric average return of the portfolio is 4.44\%, while the strategic return is 3.32\%.  The diversification return, 1.12\%, is the difference.  This represents an incremental return stemming from the volatility of the assets.  The diversification return is quite significant in this example, accounting for 25\% of the return of the portfolio.  The diversification return calculated from the approximate expression in Eq.~(\ref{eq:return}) is 1.04\%, in rough agreement with the exact value.

\begin{table}[h]
\begin{center}
\caption{A portfolio with 50\% invested in the S\&P 500 Index (Total Return) and 50\% invested in the Barclays US Long Treasury Index (Total Return) on January 1, 2000.  Return data from the Vanguard Group.}
\label{tab:portfolio}
\bigskip\begin{tabular}{|c|c|c|c|} \hline\hline
Year Ended & S\&P 500 (\%) & Long Treasuries (\%) & 50/50 Portfolio (\%) \\\hline\hline
2000 &(9.10) & 20.27 & 5.58  \\
2001 &(11.89)& 4.21  &(3.84) \\
2002 &(22.10)& 16.79 &(2.65) \\
2003 & 28.68 & 2.48  & 15.58 \\
2004 & 10.88 & 7.70  & 9.29  \\
2005 & 4.91  & 6.50  & 5.71  \\
2006 & 15.79 & 1.85  & 8.82  \\
2007 & 5.49  & 9.81  & 7.65  \\
2008 &(37.00)& 24.03 &(6.48) \\
2009 & 26.46 &(12.92)& 6.77  \\\hline\hline
$\bar r$ (\%) & 1.21  & 8.07  & 4.64 \\
$g$ (\%)      &(0.95) & 7.59  & 4.44 \\
$\sigma$ (\%) & 20.03 & 10.05 & 6.51 \\
$\sigma_{ip}^2$ $(10^{-4})$ & 117.33 & (32.69) & \\
\hline
\hline
\multicolumn{4}{|c|}{${\rm Strategic\ return}\equiv\frac{1}{2}\left(g_1+g_2\right)=3.32\%$}\\
\multicolumn{4}{|c|}{${\rm Diversification\ return}\equiv g_p-\frac{1}{2}\left(g_1+g_2\right)=1.12\%$}\\
\multicolumn{4}{|c|}{$\approx\frac{1}{4}(\sigma_1^2+\sigma_2^2-\sigma_{1p}^2-\sigma_{2p}^2)
=1.04\%$}\\\hline\hline
\end{tabular}
\end{center}
\end{table}

Another approximate expression for the diversification return of a rebalanced portfolio is obtained directly from Eq.~(\ref{eq:sigmap}):
\begin{equation}
{\rm Diversification\ Return} \approx \frac{1}{2}\left(\sum_i w_i \sigma^2_i-\sigma^2_p\right)\;.
\end{equation}
This equation is interpreted by Erb and Harvey (2006) to mean that the diversification return is the result of the reduction of the portfolio variance with respect to the weighted-average variance of the assets.  However, the true source of the diversification return is the rebalancing; the reduction of the portfolio variance is simply a consequence of diversification.  We will show in the next section that a diversified portfolio that does not rebalance, while it generally has a reduced variance, does not generate a diversification return.

Following Gorton and Rouwenhorst (2006b), we derive yet another alternative approximate expression for the diversification return of a rebalanced portfolio.  Using Eqs.~(\ref{eq:simple}) and (\ref{eq:arithmetic}), one can show that
\begin{equation}
\sigma_{ip}^2=\sum_iw_j\sigma_{ij}^2
\end{equation}
where $\sigma_{ij}^2$ is the covariance of assets $i$ and $j$ (with $\sigma_{ii}^2\equiv \sigma_i^2$).  This is related to the correlation of the assets, $\rho_{ij}$, by
\begin{equation}
\sigma_{ij}^2=\rho_{ij}\sigma_i\sigma_j\;.
\end{equation}
Putting this together with Eq.~(\ref{eq:return}), we obtain
\begin{equation}
{\rm Diversification\ Return} \approx \frac{1}{2}\left(\sum_i w_i\sigma_i^2-\sum_i\sum_j w_iw_j\rho_{ij}\sigma_i\sigma_j\right)
\label{eq:alternative}
\end{equation}

Consider the case where the assets are all perfectly correlated, $\rho_{ij}=1$ for all $i,j$.  Equation
(\ref{eq:alternative}) reduces to
\begin{equation}
{\rm Diversification\ Return} \approx \frac{1}{2}\left(\sum_i w_i\sigma_i^2-\left(\sum_i w_i\sigma_i\right)^2\right)\;\;\;\;({\rm for}\ \rho_{ij} = 1)
\label{eq:rho}
\end{equation}
that is, the diversification return is the difference between the weighted-average variance and the square of the weighted-average standard deviation of the assets. This does not vanish in general, so even a portfolio that is composed of assets that are perfectly correlated generates a diversification return.  This is not a diversified portfolio in the usual sense, where one combines assets that are not perfectly correlated with each other.  Thus even a portfolio that is not diversified in the usual sense generates a diversification return.

As an example of this last point, consider an equally-weighted portfolio of two assets with simple returns of $+25\%, -10\%$ and $+50\%, -20\%$ over two holding periods.  The simple returns of the two assets are perfectly correlated.  The strategic return of this portfolio, defined by Eq.~(\ref{eq:strategic}), is $+7.80\%$.  The geometric average return of the portfolio, rebalanced to equal weights after the first holding period, is $+8.11\%$.  The difference, $+0.31\%$, is the diversification return generated by the rebalancing.

If, in addition to being perfectly correlated, the assets had identical standard deviations, then the diversification return given by Eq.~(\ref{eq:rho}) would vanish.  This corresponds to a portfolio in which all the assets have identical returns, in which case there is no need to rebalance, and hence no opportunity to earn a diversification return.

Erb and Harvey (2006) and Idzorek (2006,2007) quote the formula
\begin{equation}
{\rm Diversification\ Return} \approx \frac{1}{2}\left(1-\frac{1}{N}\right)\overline{\sigma^2}\left(1-\overline{\rho}\right)
\end{equation}
for an equally-weighted portfolio of $N$ assets, where $\overline{\sigma^2}$ is the average variance of the $N$ assets and $\overline{\rho}$ is the average correlation of the $N(N-1)$ pairs of assets.\footnote{Idzorek (2006,2007) has the square of the average standard deviation in place of the average variance.}  This equation does not follow mathematically from Eq.~(\ref{eq:alternative}), and should not be trusted.  In particular, the diversification return does not vanish in general if $\overline{\rho}=1$, as we showed in Eq.~(\ref{eq:rho}).

Booth and Fama (1992) give many examples of diversification returns from rebalanced portfolios of stock and bond indicies.  Erb and Harvey (2006) give examples of diversification returns from rebalanced portfolios constructed from collateralized commodity futures, as well as in combination with the S\&P 500 index.  They also quote the results of a simulation involving 40 uncorrelated assets, each of which has a geometric average return of zero and a standard deviation of $30\%$.  They find that an equally-weighted, rebalanced portfolio has an average return of $4.3\%$ by running 10,000 simulations over a 45 year period.  They dub this ``turning water into wine,'' since each asset has a vanishing geometric average return.  The entire return of the portfolio is a diversification return.

We can estimate the diversification return of the Erb and Harvey simulation as follows. The geometric average return of each asset is zero, and its variance is $(30\%)^2=0.09$.  Using Eq.~(\ref{eq:alternative}), together with the fact that the assets are uncorrelated and $w_i=1/40$ yields
\begin{equation}
{\rm Diversification\ Return} \approx \frac{1}{2}\sum_iw_i(1-w_i)\sigma_i^2 \approx 4.4\%\;\;\;\;
({\rm for}\ \rho_{ij}=\delta_{ij})
\end{equation}
in rough agreement with the results of the simulation.

Swensen (2005,2009) provides examples of the incremental return generated by daily rebalancing of the Yale University investment portfolio.  He refers to this as a rebalancing bonus, since the primary reason for rebalancing is to maintain a constant risk profile.  The incremental return generated from this activity is thus regarded as an auxiliary benefit.  The rebalancing bonus is a diversification return.

\section{Buy-and-hold portfolio}\label{sec:buy}

We now turn our attention to a very different type of portfolio, in which there is no rebalancing.  We consider a buy-and-hold portfolio, in which the initial asset weights, $f_i$, are fixed ($\sum_i f_i=1$).  If the assets have geometric average returns $g_i$ over $T$ holding periods, then the portfolio geometric average return is given by
\begin{equation}
(1+g_p)^T=\sum_i f_i (1+g_i)^T\;.
\label{eq:buy}
\end{equation}
This equation makes explicit that the portfolio geometric average return is simply a function of the geometric average returns of the assets.  It has the same qualities as the strategic return of a rebalanced portfolio, given by Eq.~(\ref{eq:strategic}).  There is no dependence on the variances of the assets or their covariances with the portfolio.  The portfolio has a smaller variance than the initially-weighted-average of the assets' variances, but that alone does not imply a diversification return.

For $g_iT\ll 1$, Eq.~(\ref{eq:buy}) may be approximated by
\begin{equation}
g_p=\sum_i f_i g_i\;.
\label{eq:fixed}
\end{equation}
There is no incremental return.  In contrast, for a rebalanced portfolio, there is a diversification return, Eq.~(\ref{eq:totalreturn}), even if $g_i\ll 1$.

Consider a buy-and-hold portfolio in which the assets have identical geometric average returns.  The geometric average return of the portfolio is the same as the geometric average return of the assets, despite the fact that the variance of the portfolio is less than the initially-weighted-average variance of the assets.  This demonstrates that there is more to diversification return than just the decrease of the portfolio's variance with respect to the weighted-average variance of the assets.  The diversification return is collected only through the buying and selling of assets that occurs upon rebalancing.  Diversification is a necessary, but not sufficient, condition for a diversification return.

Nevertheless, it is possible for a buy-and-hold portfolio to generate incremental returns with respect to the initially-weighted-average asset returns given by the right-hand side of Eq.~(\ref{eq:fixed}).  This can occur because the assets that perform the best will have an increased weight in the portfolio over time, while those that underperform will have a decreased weight.  This is not a diversification return, but rather an incremental return associated with a trading strategy, namely a strategy not to trade.  Letting the asset weights drift with the performance of the assets changes the risk profile of the portfolio, in contrast with a rebalanced portfolio, which maintains a constant risk profile.

A simple example demonstrates the principle.  Consider a portfolio consisting of two assets that are initially equally weighted.  One asset has a geometric average return of $+10\%$, the other $-10\%$.  After ten holding periods the portfolio has gained $47.12\%$, corresponding to a geometric average return of $3.94\%$.  At the end of the holding period the weights of the assets have drifted from $50/50$ to $88/12$.

If each asset has a geometric average return of zero, then Eq.~(\ref{eq:buy}) shows that the portfolio has a vanishing geometric average return.  This is in contrast to a rebalanced portfolio, which may generate a nonvanishing geometric average return even if it is composed of assets with vanishing geometric average returns. This is a diversification return.  An explicit example of a two-asset portfolio was given in the previous section.  We also mentioned the ``turning water into wine'' simulation of Erb and Harvey (2006), where an equally-weighted, rebalanced portfolio of uncorrelated assets with vanishing geometric average returns yields a nonvanishing diversification return of $4.3\%$.  However, it is a mystery how this same simulation, but with an initially-equally-weighted buy-and-hold portfolio, could yield a nonvanishing return of $3.8\%$, as claimed.  Based on Eq.~(\ref{eq:buy}), the return of the buy-and-hold version of this portfolio should be zero.

Gorton and Rouwenhorst (2006b) regard the diversification return of a portfolio as the difference between the portfolio geometric average return and the (weighted) geometric average returns of the individual assets, regardless of whether the weights are constant or not.  This is the definition of diversification return for a rebalanced portfolio, as evidenced by Eq.~(\ref{eq:diversification}).  They are proposing to generalize this definition to a portfolio in which the weights are not held constant.  They do not specify how to calculate the (weighted) geometric average returns of the individual assets if the weights are not held constant.

Regardless of how one calculates this weighted average, this definition of diversification return applied to a buy-and-hold portfolio is not an appropriate generalization of the diversification return of a rebalanced portfolio, because the two types of incremental returns have completely different sources.
As discussed above, the incremental return of a buy-and-hold portfolio stems from the fact that the assets that perform the best will become a larger fraction of the portfolio.  This is completely different from the diversification return of a rebalanced portfolio, which is driven by selling high and buying low.  It is more natural to regard the return of a buy-and-hold portfolio, Eq.~(\ref{eq:buy}), as a generalization of the strategic return of a rebalanced portfolio, Eq.~(\ref{eq:strategic}).

Erb and Harvey (2006) take a different approach to a buy-and-hold portfolio.  Like Gorton and Rouwenhorst (2006b), they define the diversification return as the difference between the geometric average return of the portfolio and the (weighted) geometric average returns of the individual assets, where they specify that the weights are the average weights of the assets over the sum of the holding periods.  They break this diversification return into two pieces: a variance reduction benefit, and an impact of not rebalancing.  The variance reduction benefit is calculated from Eq.~(\ref{eq:return}), that is, it is the diversification return of a rebalanced portfolio.  The impact of not rebalancing, which they also refer to as a covariance drag, takes into account the effect of the variation of the asset weights.  The diversification return is the sum of these two pieces.

This approach is mathematically correct, as evidenced by a detailed example of a 50/50 portfolio consisting of the S\&P 500 index and heating oil futures.  However, we have argued that the diversification return of a rebalanced portfolio, Eq.~(\ref{eq:return}), is earned by rebalancing, and is not due solely to a reduction in variance.  Hence it is misleading to refer to the result of Eq.~(\ref{eq:return}) as a variance reduction benefit.

In the example of a 50/50 buy-and-hold portfolio consisting of the S\&P 500 index and heating oil futures, Erb and Harvey (2006) find a very small incremental return above that given by the right-hand side of Eq.~(\ref{eq:fixed}).  This is because the two assets have similar geometric average returns, $+6.76\%$ and $+8.21\%$, respectively.  The large incremental return of the buy-and-hold version of the Gorton and Rouwenhorst (2006a) commodity futures index stems from the fact that the assets have widely different geometric average returns.  This is discussed in the next section.

\section{The commodity return puzzle}\label{sec:commodity}

Gorton and Rouwenhorst (2006a) construct an equally-weighted collateralized commodity futures index and show that it has a significant excess return (above the risk-free rate).  Erb and Harvey (2006) show that the average excess return of the commodity futures in the index is close to zero.  They argue that the excess return of the index is mostly a diversification return, as the index rebalances monthly.  Gorton and Rouwenhorst (2006b) counter that their index produces an even higher excess return if it is not rebalanced.  Idzorek (2006,2007) refers to the dramatic difference between the average individual commodity return and an equally-weighted portfolio of commodities, and the disagreement over the importance of the different sources of return between Erb and Harvey (2006) and Gorton and Rouwenhorst (2006a,2006b), as two aspects of the {\it commodity return puzzle}.

The discussion in this paper suggests a resolution to these two puzzles.  First consider the rebalanced index.  It is plausible that the excess return of 4.52\% found by Gorton and Rouwenhorst (2006a) is indeed mostly a diversification return.  The size of the excess return is consistent with this explanation.  The average standard deviation of the commodity futures in the index is 30\%, the same as the standard deviation of the individual assets used in the Erb and Harvey (2006) ``turning water into wine'' simulation discussed above, which yielded a return of 4.3\%. That simulation used uncorrelated assets, while Idzorek (2006,2007) states that the average pair-wise correlation of the commodities in the index is 0.1, which may reduce the diversification return slightly.  Even taking this into account, most of the excess return of the Gorton and Rouwenhorst index appears to be a diversification return, and is large because the assets from which the index is constructed are so volatile.  This is the conclusion reached by Erb and Harvey (2006).

Now consider the unrebalanced index, which is a buy-and-hold portfolio.  As we discussed in Section~\ref{sec:buy}, a buy-and-hold portfolio can generate an incremental return with respect to the initially-weighted-average return of the assets if the assets have widely different geometric average returns.  This is indeed the case for the commodity futures in the Gorton and Rouwenhorst index (although it is not the case for the Erb and Harvey ``turning water into wine'' simulation, where each asset has a vanishing geometric average return.).  Although the equally-weighted geometric average return of the commodity futures is close to zero, the geometric average returns of the individual commodity futures are widely varying, and yield a significant incremental return.  We argued in Section~\ref{sec:buy} that this should not be regarded as a diversification return, but rather as a generalization of a strategic return.

Thus there is no contradiction.  Both the rebalanced index and the unrebalanced index yield a significant excess return, but for totally different reasons.  The excess return of the rebalanced index is mostly a diversification return, which is driven by the selling of assets that have appreciated in relative value and the buying of assets that have declined in relative value, as discussed in Section~\ref{sec:source}.  In contrast, the excess return of the unrebalanced index is driven by the fact that the assets that perform the best will become a larger fraction of the portfolio, while those that perform poorly will become a smaller fraction, as discussed in Section~\ref{sec:buy}.  This changes the risk profile of the portfolio, unlike the rebalanced portfolio, which maintains a constant risk profile.

The Gorton and Rouwenhorst index is not strictly an equally-weighted, monthly rebalanced index because the index begins with 9 commodity futures in 1959 and ends up with 36 commodity futures by 2003.  However, the commodity futures are added one by one to the index,\footnote{The only exceptions are Palladium and Zinc, which are added to the index simultaneously.} and each month that one is added the index is rebalanced to equal weights, just as it is in all other months.  This is the natural generalization of an equally-weighted, monthly rebalanced index when securities are gradually added to the index.  The monthly rebalancing activity generates a diversification return, as discussed in Section~\ref{sec:source}.  In the buy-and-hold version of the index, when the $N^{\rm th}$ commodity future becomes available to add to the index, a fraction $1/N$ of the index is sold and the proceeds are reinvested in the $N^{\rm th}$ commodity future, without rebalancing the relative weights of the other $N-1$ commodity futures.  This is the natural generalization of a buy-and-hold portfolio when securities are gradually added to the portfolio.  The commodity futures that perform the best will become a larger fraction of the portfolio, while those that perform poorly will become a smaller fraction.  This generates an incremental return in the same way as in the case of a true buy-and-hold portfolio, as discussed in Section~\ref{sec:buy}.

\newpage

\section{Conclusions}\label{sec:conclusions}

The diversification return of a rebalanced portfolio is often ascribed to a reduction in variance.  We have argued that the underlying source of the diversification return is the rebalancing, which forces the investor to sell assets that have appreciated in relative value and buy assets that have declined in relative value, as measured by their weights in the portfolio.  A buy-and-hold portfolio, while it generally has a lower variance than the weighted-average variance of its constituents, does not earn a diversification return. However, a buy-and-hold portfolio can benefit from the fact that the assets that perform the best become a larger fraction of the portfolio over time.  This changes the risk profile of the portfolio, unlike a rebalanced portfolio, which maintains a constant risk profile.

These results resolve two of the three aspects of the commodity return puzzle, articulated by Idzorek (2006,2007).  The commodity futures index of Gorton and Rouwenhorst (2006a), which is rebalanced monthly, earns a significant diversification return because the underlying commodity futures are so volatile, as argued by Erb and Harvey (2006).  In contrast, the buy-and-hold version of this same index earns a significant return because the underlying commodity futures have a wide range of geometric average returns.  Thus we are able to explain the source of the return for both the rebalanced and the buy-and-hold versions of the index, and also to explain why there is no contradiction between the returns of these two versions of the index.  The third aspect of the commodity return puzzle, the unexplained historical return premium of individual commodity futures, is not addressed in this paper.

Diversification is often described as the only ``free lunch'' in finance, as it allows for the reduction of risk for a given expected return.  Diversification return might be described as the ``free dessert,'' as it is an incremental return earned while maintaining a constant risk profile.  However, it is necessary to perform the contrarian activity of rebalancing in order to earn the diversification return; diversification is a necessary, but not sufficient, condition. While an unrebalanced portfolio generally has reduced risk, it does not earn a diversification return, and also suffers from a varying risk profile.  The control of risk, in combination with the diversification return, is a powerful argument in favor of rebalanced portfolios.

The excess return (over the risk-free rate) of the Gorton and Rouwenhorst (2006a) commodity futures index is consistent with a diversification return.  By extension, the excess return of other rebalanced commodity futures indices may be largely due to a diversification return.  The buy-and-hold version of the Gorton and Rouwenhorst index also earns an excess return, but for a totally different reason: some commodity futures have produced such large gains that they more than compensate for the commodity futures that have done poorly.  This approach to investing is similar to that of venture capital, where a few big winners more than compensate for the many losers.

\section*{Acknowledgements}

I am grateful for correspondence with Tom Idzorek and for assistance from Carissa Holler.

\newpage

\section*{Appendix}\label{sec:appendixA}

By Taylor expanding the relation between the compound return and the simple return, $r$, about the arithmetic average return, $\bar r$, and then averaging, Booth and Fama (1992) derive the relation
\begin{equation}
C = \ln(1+\bar r) - \frac{1}{2}\frac{\sigma^2}{(1+\bar r)^2}+\cdots
\label{eq:geometricexpansion}
\end{equation}
where $C$ is the (average) compound return and $\sigma^2$ is the variance of the simple returns.  Replacing the compound return with the geometric average return, $g$, via $C=\ln(1+g)$, and exponentiating both sides of the equation yields
\begin{equation}
1+g\approx (1+\bar r)e^{ - \frac{1}{2}\frac{\sigma^2}{(1+\bar r)^2}}
\end{equation}
where we have dropped the higher terms in the expansion.  Expanding the exponential
and keeping the leading terms in powers of $\bar r$ yields Eq.~(\ref{eq:geometric}).

\newpage

\section*{References}\label{sec:references}


Booth, David G., and Eugene F.~Fama. 1992. ``Diversification Returns and Asset Contributions.''
{\it Financial Analysts Journal}, vol.~48, no.~3 (May/June): 26-32.
\medskip

\noindent Erb, Claude B., and Campbell R.~Harvey. 2006. ``The Tactical and Strategic Value of Commodity Futures.''
{\it Financial Analysts Journal}, vol.~62, no.~2 (March/April): 69-97.
\medskip

\noindent Gorton, Gary B., and Rouwenhorst, K.~Geert. 2006a. ``Facts and Fantasies About Commodity Futures.''
{\it Financial Analysts Journal}, vol.~62, no.~2 (March/April): 47-68.
\medskip

\noindent Gorton, Gary B., and Rouwenhorst, K.~Geert. 2006b. ``A Note on Erb and Harvey (2005).'' Yale ICF Working Paper No.~06-02. http://papers.ssrn.com/sol3/papers.cfm?abstract\_id=869064.\\

\noindent Idzorek, Thomas M. 2006. ``Strategic Asset Allocation and Commodities,'' Ibbotson Associates.
\medskip

\noindent Idzorek, Thomas M. 2007. ``Commodities and Strategic Asset Allocation,'' in {\it Intelligent Commodity Investing}, edited by Hilary Till and Joseph Eagleeye. London: Risk Books.
\medskip

\noindent Swensen, David F. 2005. {\it Unconventional Success}. New York: Free Press. p.~197.
\medskip

\noindent Swensen, David F. 2009. {\it Pioneering Portfolio Management}. New York: Free Press. p.~135.
\medskip


\end{document}